\documentclass[runningheads]{llncs}
\usepackage[T1]{fontenc}
\usepackage{graphicx}
\usepackage{amsmath}
\usepackage{amsfonts}
\usepackage{array}
\usepackage{hhline}
\usepackage[dvipsnames]{xcolor}
\usepackage{multirow}
\usepackage{url}

\usepackage{soul}

\usepackage[colorlinks=true]{hyperref}
\begin{document}
\title{Generating Realistic Brain MRIs via a Conditional Diffusion Probabilistic Model}
%
\titlerunning{Conditional DPM for 3D MRI Generation}
%

\author{Wei Peng\inst{1}\orcidID{0000-0002-2892-5764} \and
Ehsan Adeli\inst{1}\orcidID{[0000-0002-0579-7763} \and
Tomas Bosschieter \inst{1}\orcidID{0000-0001-9726-6400}\and
Sang Hyun Park \inst{2}\orcidID{0000-0001-7476-1046}
 \and
Qingyu Zhao\inst{1}\orcidID{0000-0002-6368-0889} \and
Kilian M. Pohl\inst{1}\thanks{Corresponding Author: kpohl@stanford.edu}\orcidID{0000-0001-5416-5159}}
\authorrunning{Peng et al.}

\institute{{\small Stanford University, Stanford, CA 94305} \and
 Daegu Gyeongbuk Institute of Science \& Technology, South Korea
}

\maketitle              
%
\begin{abstract}
As acquiring MRIs is expensive, neuroscience studies struggle to attain a sufficient number of them for properly training deep learning models. This challenge could be reduced by MRI synthesis, for which Generative Adversarial Networks (GANs) are popular. GANs, however, are commonly unstable and struggle with creating diverse and high-quality data. A more stable alternative is Diffusion Probabilistic Models (DPMs) with a fine-grained training strategy. To overcome their need for extensive computational resources, we propose a conditional DPM (cDPM) with a memory-efficient process that generates realistic-looking brain MRIs. To this end, we train a 2D cDPM to generate an MRI subvolume conditioned on another subset of slices from the same MRI. By generating slices using arbitrary combinations between condition and target slices, the model only requires limited computational resources to learn interdependencies between slices even if they are spatially far apart. After having learned these dependencies via an attention network, a new anatomy-consistent 3D brain MRI is generated by repeatedly applying the cDPM. Our experiments demonstrate that our method can generate high-quality 3D MRIs that share a similar distribution to real MRIs while still diversifying the training set. The code is available at { \url{https://github.com/xiaoiker/mask3DMRI_diffusion}} and also will be released as part of MONAI, at {\url{https://github.com/Project-MONAI/GenerativeModels}}. 
\end{abstract}

\section{Introduction}
The synthesis of medical images has great potential in aiding tasks like improving image quality, imputing missing modalities \cite{zheng2022diffusion}, performing counterfactual analysis~\cite{pawlowski2020deep}, and modeling disease progression~\cite{zhao2021longitudinal,jung2021conditional,jung2023conditional}.  However, synthesizing brain MRIs is non-trivial as they are of high dimension, yet the training data are relatively small in size (compared to 2D natural images). High-quality synthetic MRIs have been produced by conditional models based on real MRI of the same subject acquired with different MRI sequences~\cite{shin2018medical,MultiContrastGAN2019,yu20183d,ouyang2021representation}. However, such models require large data sets (which are difficult to get) and fail to significantly improve data diversity~\cite{karras2019style,xing2021cycle}, i.e., producing MRIs substantially deviating from those in the training data; data diversity is essential to the generalizability of large-scale models~\cite{xing2021cycle}.
Unconditional models based on Generative Adversarial Networks (GANs) bypass this drawback by generating new, independent MRIs from random noise~\cite{bermudez2018learning,han2018gan}. However, these models often produce lower quality MRIs as they currently can only be trained on lower resolution MRIs or 2D slices due to their computational needs ~\cite{gulrajani2017improved,kwon2019generation}. Furthermore, GAN-based models are known to be unstable during training and even can suffer from mode collapse~\cite{goodfellow2020generative}. An alternative is diffusion probabilistic models (DPMs)~\cite{ho2020denoising,sohl2015deep}, which formulate the fine-grained mapping between data distribution and Gaussian noise as a gradual process modeled within a Markov chain. Due to their multi-step, fine-grained training strategy, DPMs tend to be more stable during training than GANs and therefore are more accurate for certain medical imaging applications, such as segmentation and anomaly detection~\cite{wolleb2022diffusion,la2022anatomically}. However, DPMs tend to be computationally too expensive to synthesize brain MRI at full image resolution \cite{pinaya2022brain,dorjsembe2022three}. We address this issue by proposing a memory-efficient 2D conditional DPM (cDPM) that relies on learning the interdependencies between 2D slices to produce high-quality 3D MRI volumes. 

\begin{figure}[!t]
\centering
\includegraphics[width=1\columnwidth]{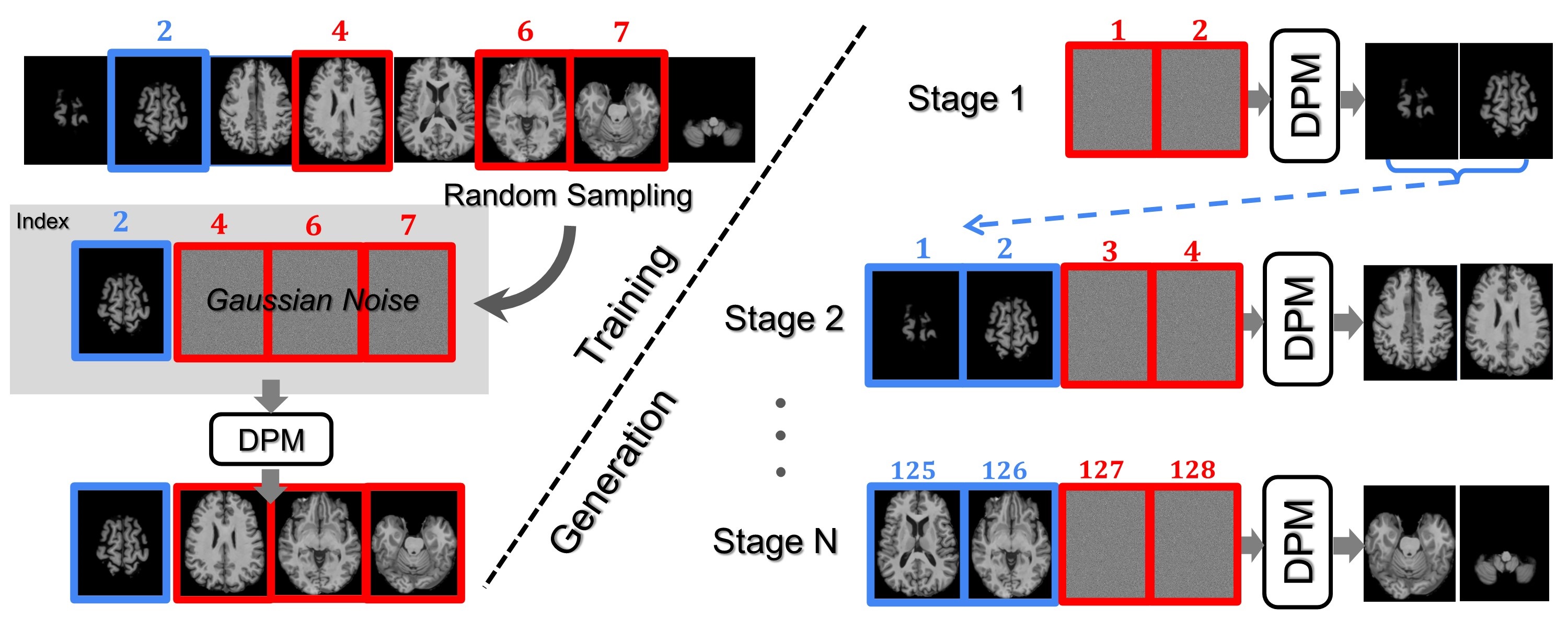}
\caption{A memory efficient DPM. \textbf{Left:} Based on \textcolor{cyan}{`condition' slices}, cDPM learns to generate \textcolor{red}{`target' slices}. \textbf{Right:} A new 3D MRI is created by repeatedly running the trained model to synthesize target slices conditioned on those it created in prior stages.} 
\label{overview}
\end{figure}

Unlike the sequence of 2D images defining a video, all 2D slices of an MRI are interconnected with each other as they define a 3D volume capturing brain anatomy. Our cDPM learns these interdependencies (even between distant slices) by training an attention network~\cite{vaswani2017attention} on arbitrary combinations of condition and target slices. Once learned, the cDPM creates new samples while capturing brain anatomy in 3D. It does so by producing the first few slices from random noise and then using those slices to synthesize subsequent ones (see Fig.~\ref{overview}). We show that this computationally efficient conditional DPM can produce MRIs that are more realistic than those produced by GAN-based architectures. Furthermore, our experiments reveal that cDPM is able to generate synthetic MRIs, whose distribution matches that of the training data. 

\section{Methodology}
We first review the basic DPM framework for data generation (Section~\ref{sec31}). Then, we introduce our efficient strategy for generating 3D MRI slices (Section \ref{sec33}) and finally describe the neural architecture of cDPMs (Section \ref{sec34}).

\subsection{Diffusion Probabilistic Model}
\label{sec31}
The Diffusion Probabilistic Model (DPM)~\cite{sohl2015deep,ho2020denoising} generates MRIs from random noise by iterating between mapping 1) data gradually to noise (a.k.a., Forward Diffusion Process) and 2) noise back to data (a.k.a., Reverse Diffusion Process).

\subsubsection{Forward Diffusion Process (FDP)}
Let real data $X_0 \sim \textbf{q}$ sampled from the (real data) distribution $\textbf{q}$ be the input to the FDP. FDP then simulates the diffusion process that turns $X_0$ after $T$ perturbations into Gaussian noise $X_T \sim \mathcal{N}(\textit{0}, \textit{I}\,)$, where $\mathcal{N}$ is the Gaussian distribution with zero mean and the variance being the identity matrix $\textit{I}$. This process is formulated as a Markov chain,
whose transition kernel $q(X_t|X_{t-1})$ at time step $t \in \left\{0,\ldots, T\right\}$ is defined as 
\begin{equation}
     q(X_t|X_{t-1}) := \mathcal{N}(X_t;\sqrt{1-\beta_t}\cdot X_{t-1}, \beta_t\cdot I).
\end{equation}
The weight $\beta_t \in (0, 1)$ is changed so that the chain gradually enforces drift, i.e., adds Gaussian noise to the data. Let $\alpha_{t} := 1-\beta_t$ and $\bar{\alpha}_t := \prod_{s=1}^{t}(1-\beta_t)$, then $X_t$ is a sample of the distribution conditioned on $X_0$ as 
\begin{equation}
\label{eq:eq2}
    q(X_t|X_0) := \mathcal{N}(X_t;\sqrt{\bar{\alpha_t}}\cdot X_{0},  (1-\bar{\alpha}_t)\cdot I).
\end{equation}
Given this closed-form solution, we can sample $X_t$ at any arbitrary time step $t$ without needing to iterate through the entire Markov chain. 


\subsubsection{Reverse Diffusion Process (RDP)}
The RDP aims to generate realistic data from random noise $X_T$ by approximating the posterior distribution $p(X_{t-1}|X_t)$. It does so by going through the entire Markov chain from time step $T$ to 0, i.e.,   
\begin{equation}
p(X_{0:T}) :=  p(X_{T}) \prod_{t=1}^{T} p_{\theta}(X_{t-1}|X_t).
\label{eq:unconditionalP}
\end{equation}
Defining the conditional distribution $p_{\theta}(X_{t-1}|X_t):= \mathcal{N}(X_{t-1}; \mu_{\theta}(X_t, t), \Sigma)$ with fixed variance $\Sigma$, then (according to \cite{ho2020denoising}) the mean can be rewritten as 
\begin{equation}
\mu_{\theta}(X_t, t) = \frac{1}{\sqrt{\alpha_t}} \left( X_t - \frac{\beta_t}{\sqrt{(1-\bar{\alpha}_t)}}\epsilon_{\theta}(X_t,t) \right),
\label{eqn:mu_estimate}
\end{equation}
with $\epsilon_{\theta}(\cdot)$ being the estimate of a neural network defined by parameters $\theta$. $\theta$ minimizes the reconstructing loss defined by the following expected value 
\begin{equation*}
\mathbb{E}_{X_0 \sim \textbf{q}, t \in [0,\ldots, T], \epsilon \sim \mathcal{N}(0,I)}\left[||\epsilon - \epsilon_{\theta}(X_t,t)||_2^2\right],
\end{equation*}
where $||\cdot||_2$ is the L2 norm, and $X_t$ is inferred  from Eq. \eqref{eq:eq2} based on $X_0$. 



\subsection{Conditional Generation with DPM (cDPM)}
\label{sec33}

\noindent To synthetically create high-resolution 3D MRI, we propose an efficient cDPM model that learns the interdependencies between 2D slices of an MRI so that it can generate slices based on another set of already synthesized ones (see Fig.~\ref{overview}). 

Specifically, given an MRI $X \in \mathbb{R}^{D \times H \times W}$, we randomly sample two sets of slice indexes: the condition set $\mathcal{C}$ and the target set $\mathcal{P}$. Let $\text{len}(\cdot)$ be the number of slices in a set, then the `condition' slices are defined as $ X^{\mathcal{C}} \in \mathbb{R}^{\text{len}(\mathcal{C})\times H \times W}$ and the `target' slices as $ X^{\mathcal{P}} \in \mathbb{R}^{\text{len}(\mathcal{P})\times H \times W}$ with  $\text{len}(\mathcal{P}) \geq 1$. Confining the FDP of Section \ref{sec31} just to the target $X^{\mathcal{P}}$, the RDP now aims to reconstruct $X^{\mathcal{P}}_{t}$ for each time $t=T,T-1, \ldots, 0$ starting from random noise at $t=T$ and conditioned on $X^{\mathcal{C}}$. Let $\widetilde{X}_t$ be the subvolume consisting of $X^{\mathcal{P}}_{t}$ and $X^{\mathcal{C}}$, then the joint distribution of the Markov chain defined by Eq.~\eqref{eq:unconditionalP} now reads  
\begin{equation}
    p(X^{\mathcal{P}}_{0:T}) :=  p(X^{\mathcal{P}}_{T}) \prod_{t=1}^{T} p_{\theta}(X^{\mathcal{P}}_{t-1}|\widetilde{X}_t). 
\label{eqn:conditionalP}
\end{equation}
Observe that Eq.~\eqref{eqn:conditionalP} is equal to Eq.~\eqref{eq:unconditionalP} in case $\text{len}(\mathcal{C})=0$.

To estimate $\mu_{\theta}(\widetilde{X}_t, t)$ as described in Eq.~\eqref{eqn:mu_estimate}, we sample arbitrary index sets $\mathcal{C}$ and $\mathcal{P}$ so that $\text{len}(\mathcal{C}) + \text{len}(\mathcal{P}) \leq \tau_{\text{max}}$, where $\tau_{\text{max}}$ is the maximum number of slices based on the available resources. We then capture the dependencies across slices by feeding the index sets $\mathcal{C}$ and $\mathcal{P}$ and the corresponding slices (i.e., $X^{\mathcal{C}}$ and $X_{t}^{\mathcal{P}}$ built from $X_0 \sim \textbf{q}$) into an attention network~\cite{shaw2018self}. The neural network aims to minimize the canonical loss function 
\begin{equation}
    \text{Loss}(\theta) := \mathbb{E}_{X_0 \sim \textbf{q}, \epsilon \sim \mathcal{N}(0,I), \mathcal{C} + \mathcal{P}\leq \tau_{\text{max}},t }\left[ ||\epsilon - \epsilon_{\theta}(X_{t}^{\mathcal{P}}, X^{\mathcal{C}}, \mathcal{C},\mathcal{P}, t)||_2^2 \right]. 
\end{equation}
\begin{figure}[!t]
\centering
\includegraphics[width=1\textwidth]{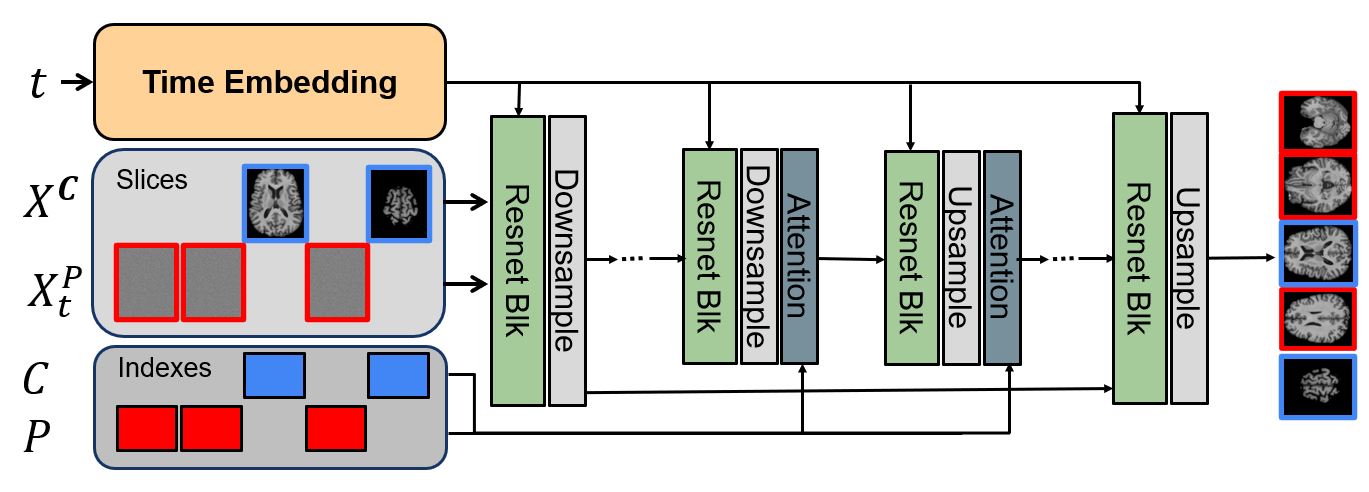}
\caption{The architecture of cDPM is a U-shape neural network with skip connections and the input at step `$t$' are slice indexes $\{\mathcal{C},\mathcal{P}\}$, condition sub-volume $X^\mathcal{C}$, and current target sub-volume $X_t^\mathcal{P}$ .} 
\label{architecture}
\end{figure}
As the neural network can now be trained on many different (arbitrary) slice combinations (defined by $\mathcal{C}$ and $\mathcal{P}$), the cDPM only requires a relatively small number of MRIs for training. Furthermore, it will learn short- and long-range dependencies across slices as the spatial distance between slices from $\mathcal{C}$ and $\mathcal{P}$ varies. Learning these dependencies (after being trained for a sufficiently large number of iterations) enables cDPMs to produce 2D slices that, when put together, result in realistic looking, high-resolution 3D MRIs.

\subsection{Network Architecture}
\label{sec34}
As done by~\cite{ho2020denoising}, cDPMs are implemented as a U-Net~\cite{ronneberger2015u} with a time embedding module (see Fig.~\ref{architecture}). 
We add a multi-head self-attention mechanism~\cite{vaswani2017attention} to model the relationship between slices. After training the cDPM as in Fig.~\ref{overview}, a 3D MRI is generated in $N$ stages. Specifically, the cDPM produces the initial set of slices of that MRI volume from random noise (i.e., unconditioned). Conditioned on those synthetic slices, the cDPM then runs again to produce a new set of slices. The process of synthetically creating slices based on ones generated during prior stages is repeated until an entire 3D MRI is produced. 


\section{Experiments}\label{sec:experiments}

\subsection{Data}
We use 1262 t1-weighted brain MRIs of subjects from three different datasets: the Alzheimer's Disease Neuroimaging Initiative (ADNI-1), UCSF (PI: V. Valcour), and SRI International (PI: E.V. Sullivan and A. Pfefferbaum)~\cite{zhang2022multi}. Processing includes denoising, bias field correction, skull stripping, and affine registration to a template, and normalizing intensity values between 0 and 1. In addition, we padded and resized the MRIs to have dimensions $128 \times 128 \times 128$ resulting in a voxel resolution of 1.375mm x 1.375mm x 1.0 mm. Splitting the MRI along the axial direction results in 2D slices. Note, this could have also been done along the sagittal or coronal direction.  
%
\begin{figure}[t!]
\centering
\includegraphics[width=1.\textwidth]{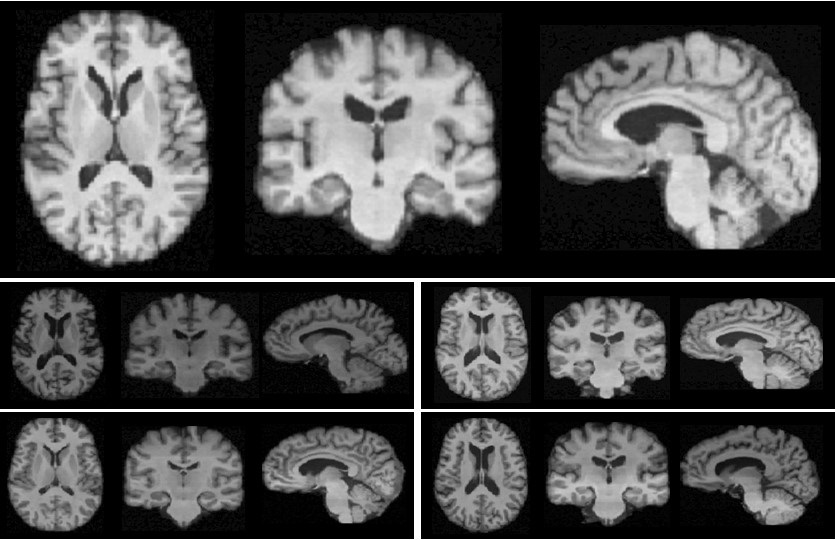}
\caption{5 MRIs generated by our conditional DPM visualized in the axial, coronal, and sagittal plane. The example in the first row is enlarged to highlight the high quality of synthetic MRIs generated by our approach. } 
\label{showcases}
\end{figure}
\begin{figure}[t!]
\centering
\begin{tabular}{c@{}c@{}c@{}c@{}c@{}c@{}c}
\includegraphics[width=0.1425\textwidth, height=0.18\textwidth]{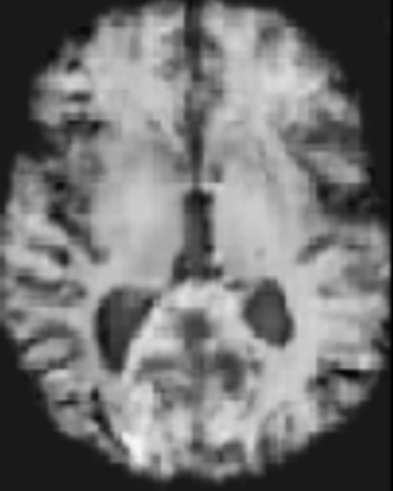}
 & \includegraphics[width=0.1425\textwidth, height=0.18\textwidth]{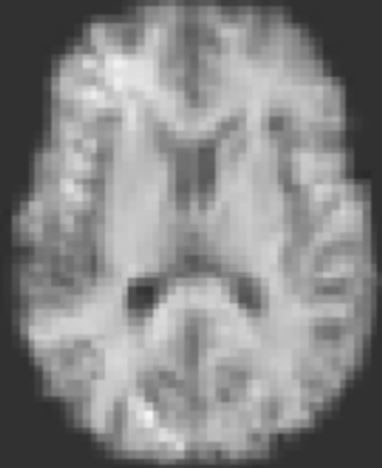}
& \includegraphics[width=0.1425\textwidth, height=0.18\textwidth]{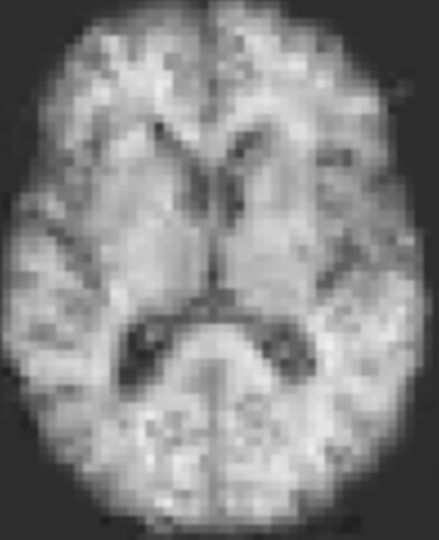} 
& \includegraphics[width=0.1425\textwidth, height=0.18\textwidth]{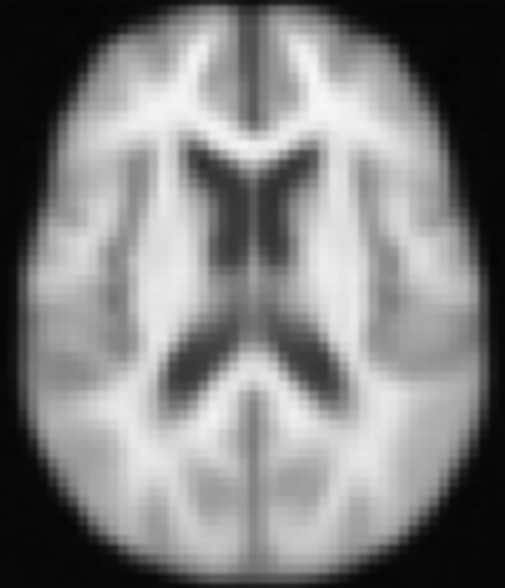}
& \includegraphics[width=0.1425\textwidth, height=0.18\textwidth]{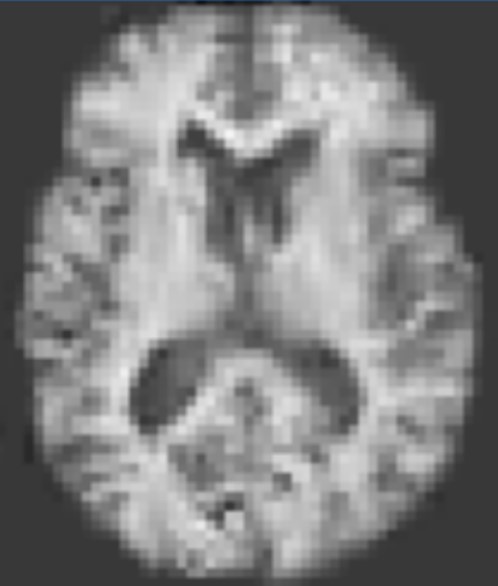}
&\includegraphics[width=0.1425\textwidth, height=0.18\textwidth]{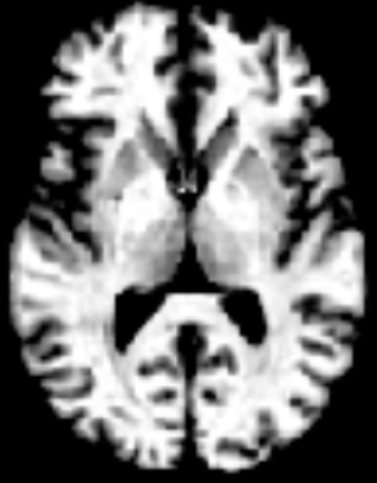}
& \includegraphics[width=0.1425\textwidth, height=0.18\textwidth]{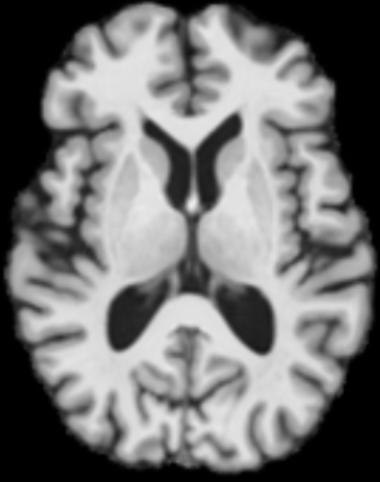}\\

\includegraphics[width=0.1425\textwidth, height=0.15\textwidth]{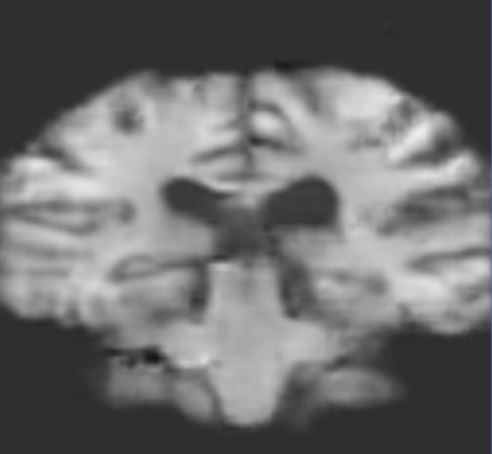}
 &\includegraphics[width=0.1425\textwidth, height=0.15\textwidth]{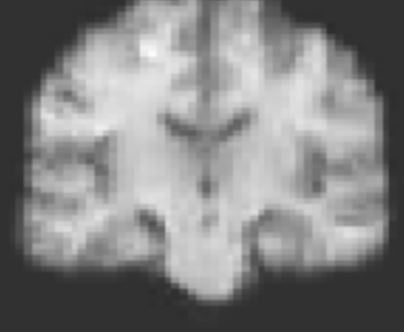}
& \includegraphics[width=0.1425\textwidth, height=0.15\textwidth]{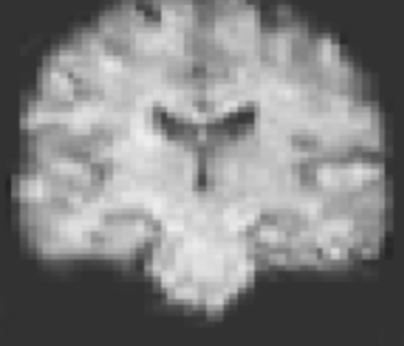} 
& \includegraphics[width=0.1425\textwidth, height=0.15\textwidth]{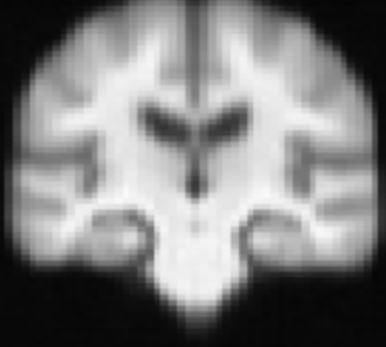}
& \includegraphics[width=0.1425\textwidth, height=0.15\textwidth]{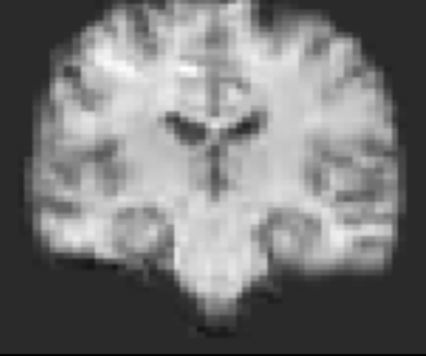}
& \includegraphics[width=0.1425\textwidth, height=0.15\textwidth]{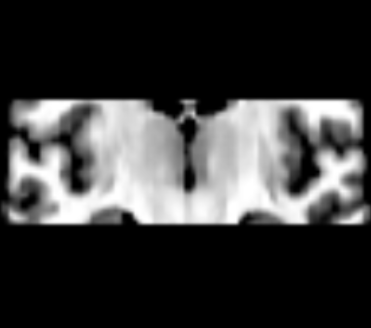}
& \includegraphics[width=0.1425\textwidth, height=0.15\textwidth]{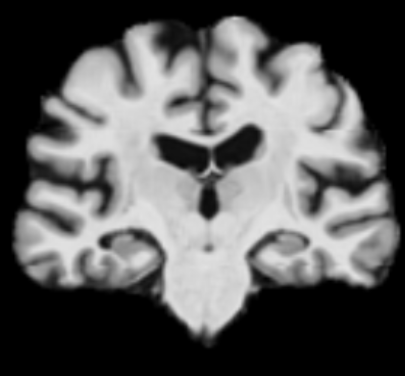}\\

\includegraphics[width=0.1425\textwidth, height=0.11\textwidth]{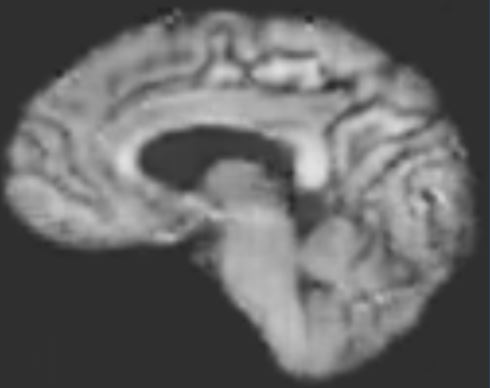}
 &\includegraphics[width=0.1425\textwidth, height=0.11\textwidth]{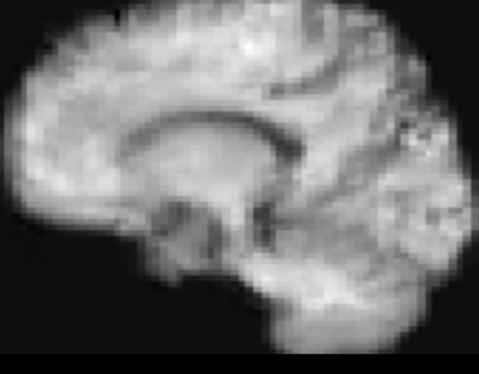}
& \includegraphics[width=0.1425\textwidth, height=0.11\textwidth]{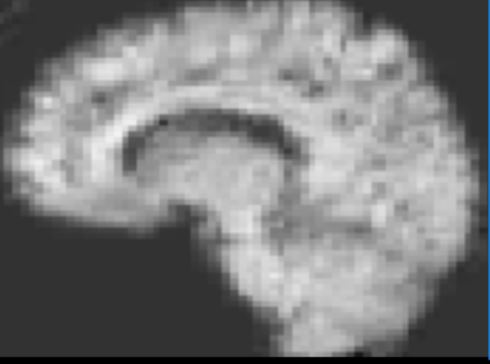} 
& \includegraphics[width=0.1425\textwidth, height=0.11\textwidth]{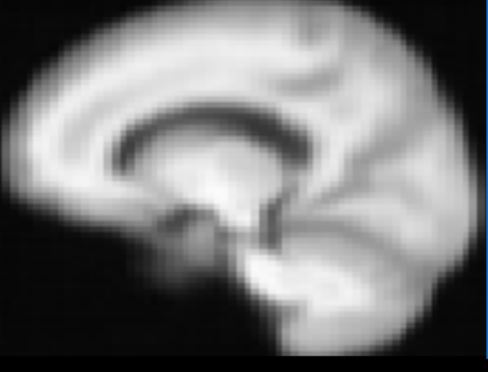}
& \includegraphics[width=0.1425\textwidth, height=0.11\textwidth]{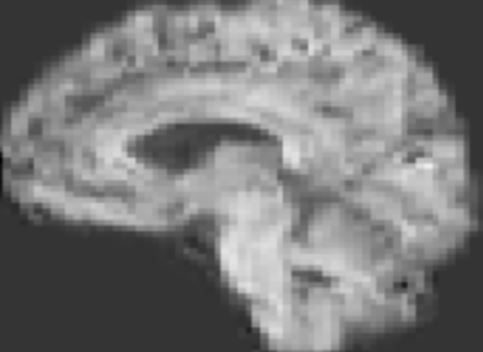}
& \includegraphics[width=0.1425\textwidth, height=0.11\textwidth]{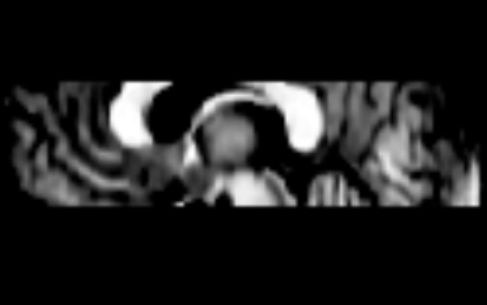}
& \includegraphics[width=0.1425\textwidth, height=0.11\textwidth]{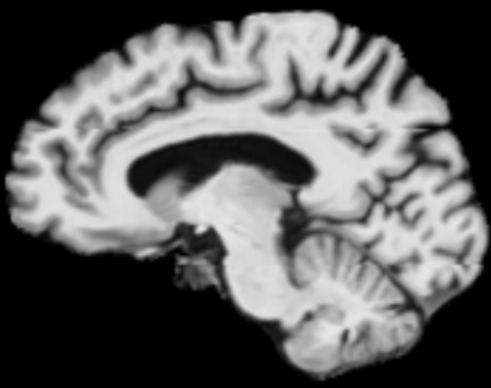}\\

\scriptsize{HA-GAN}& \scriptsize{CCE-GAN} & \scriptsize{3D-GAN-GP} & \scriptsize{3D-VAE-GAN} & \scriptsize{3D-$\alpha$-WGAN} &\scriptsize{3D-DPM}& \textbf{\scriptsize{Ours(cDPM)}} \\
\end{tabular}
\caption{3 views of MRIs generated by 7 models. Compared to the MRIs produced by the other approaches, our cDPM model generates the most realistic MRI scans that provide more distinct gray matter boundaries and greater anatomical details. }
\label{fig:comparetoGANs}
\end{figure}
\subsection{Implementation Details}
Our experiments are conducted on an NVIDIA A100 GPU using the PyTorch framework. The model is trained using 200,000 iterations with the AdamW optimizer adopting a learning rate of $10^{-4}$ and a batch size of 3. $\tau_{\textnormal{max}}$ is set to 20. After the training, cDPM generates a synthetic MRI consisting of 128 slices by following the process outlined in Fig.~\ref{overview} in N=13 stages. Each stage generates 10 slices starting with pure noise ($X^{\mathcal{C}}=\emptyset$) and (after the first stage) being conditioned on the 10 slices produced by the prior stage. After training on all real MRIs, we use the resulting conditional DPM to generate 500 synthetic MRIs.

\subsection{Quantitative Comparison}
 We evaluate the quality of synthetic MRIs based on 3 metrics: (i) computing the distance between synthetic and 500 randomly selected real MRIs via the Maximum-Mean Discrepancy (MMD) score~\cite{gretton2012kernel}, (ii) measuring the diversity of the synthetic MRIs via the pair-wise multi-scale Structure Similarity (MS-SSIM)~\cite{kwon2019generation}, and (iii) comparing the distributions of synthetic to real MRIs with respect to the 3 views via the Fr\`{e}chet Inception Distance (FID)~\cite{xing2021cycle} (a.k.a, FID-Axial, FID-Coronal, FID-Sagittal). 

We compare those scores to ones produced by six recently published methods: (i) 3D-DPM~\cite{dorjsembe2022three}, (ii) 3D-VAE-GAN~\cite{larsen2016autoencoding}, (iii) 3D-GAN-GP~\cite{gulrajani2017improved}, (iv) 3D-$\alpha$-WGAN~\cite{kwon2019generation}, (v) CCE-GAN~\cite{xing2021cycle}, and (vi) HA-GAN~\cite{sun2022hierarchical}. We needed to re-implement the first 5 methods and used the open-source code available for HA-GAN. 3D-DPM was only able to generate 32 slices at a time (due to GPU limitations) so that we computed its quality metrics by also cropping the corresponding real MRI to those 32 slices. 



\begin{figure}[!t]
\centering
\begin{tabular}{ccccc}
\includegraphics[width=0.15\textwidth, height=0.18\textwidth]{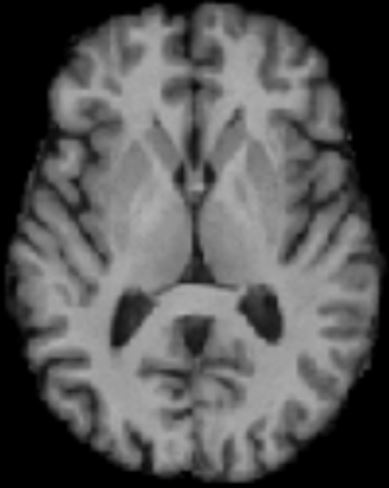}
& \includegraphics[width=0.17\textwidth, height=0.18\textwidth]{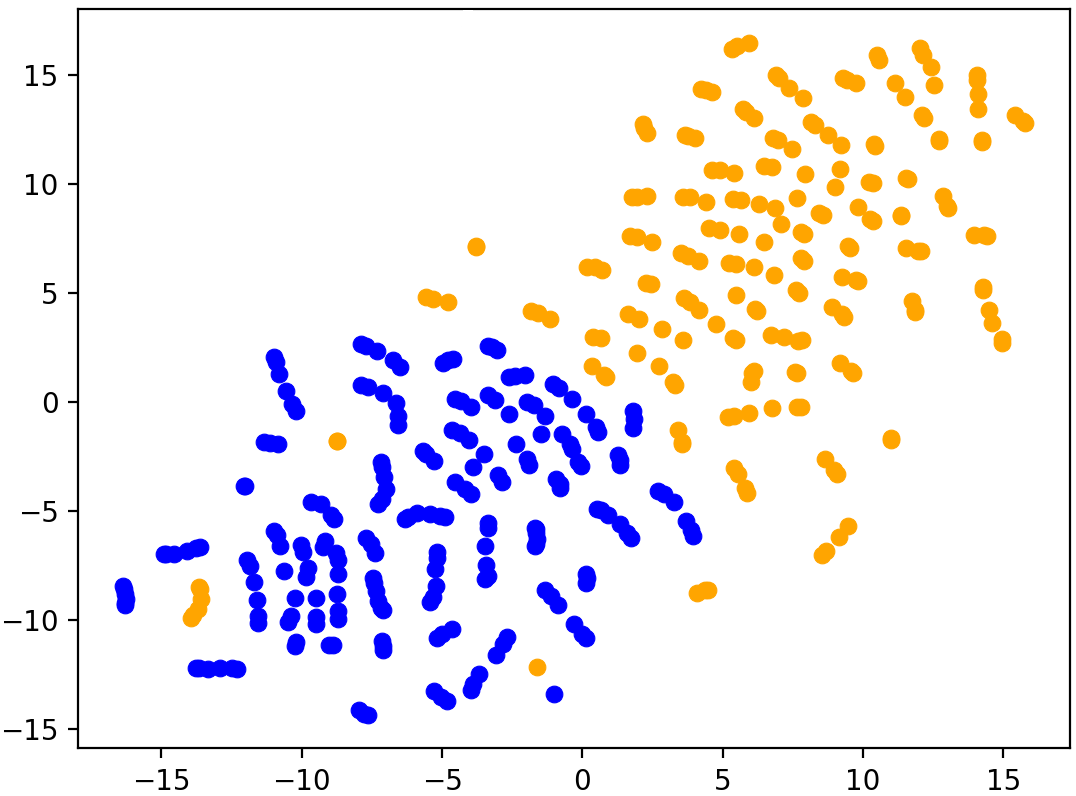}
& \includegraphics[width=0.17\textwidth, height=0.18\textwidth]{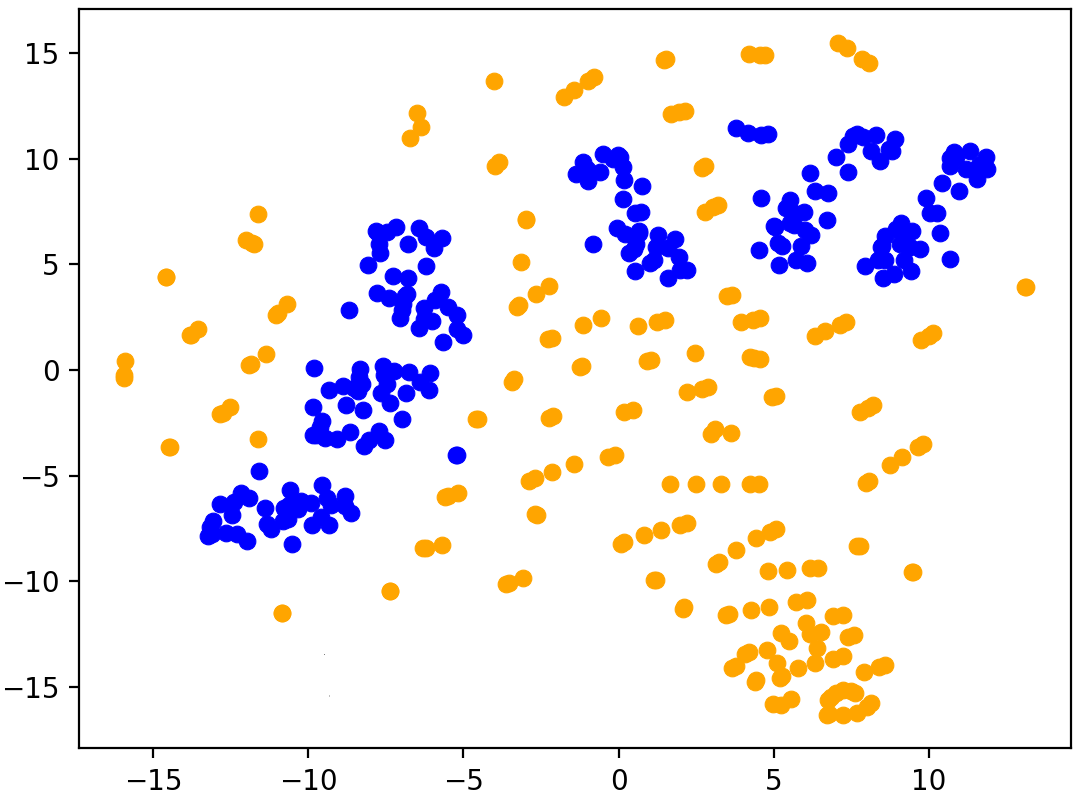}
& \includegraphics[width=0.17\textwidth, height=0.18\textwidth]{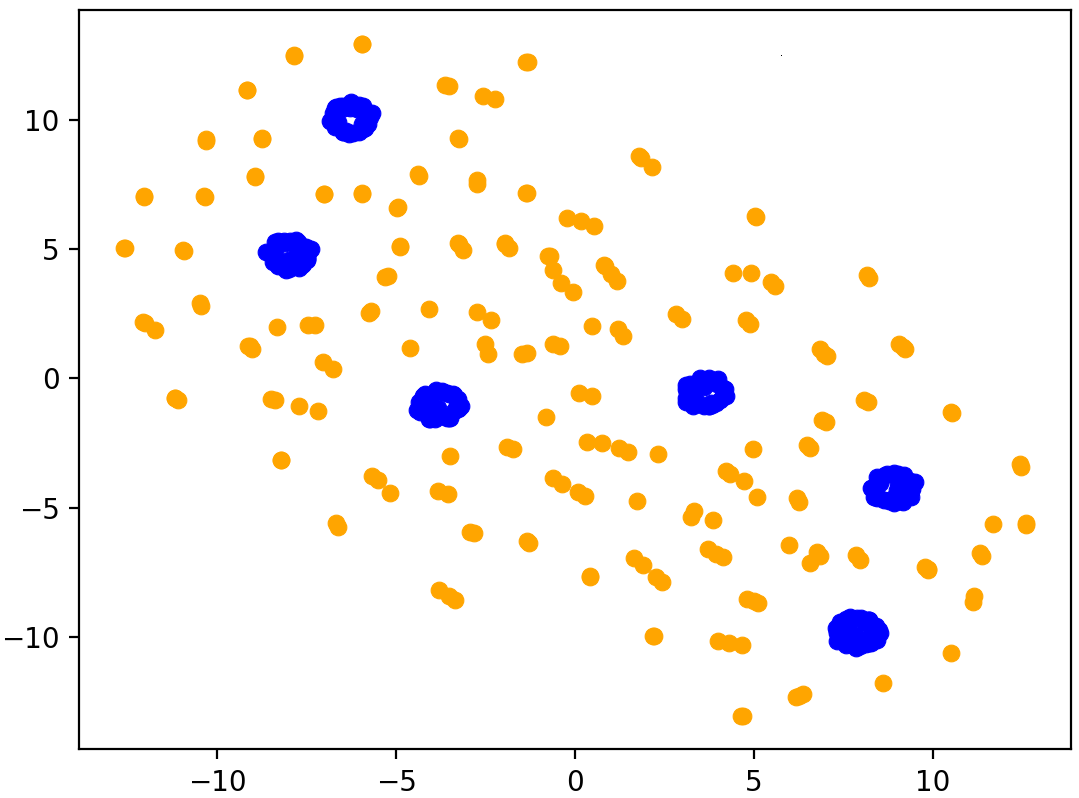}\vspace{0mm}
&\\
 \scriptsize{cDPM} &  \scriptsize{CCE-GAN} & \scriptsize{3D-GAN-GP} & \scriptsize{3D-VAE-GAN} &\vspace{0mm}\\\\
 \includegraphics[width=0.15\textwidth, height=0.18\textwidth]{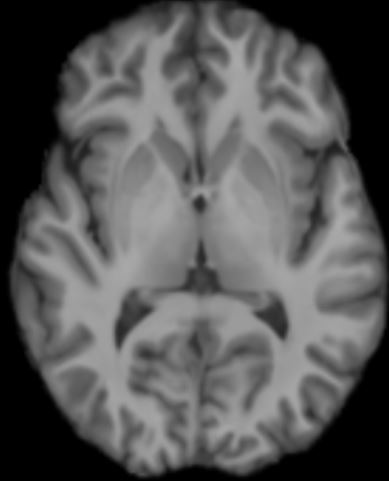}
&  \includegraphics[width=0.17\textwidth, height=0.18\textwidth]{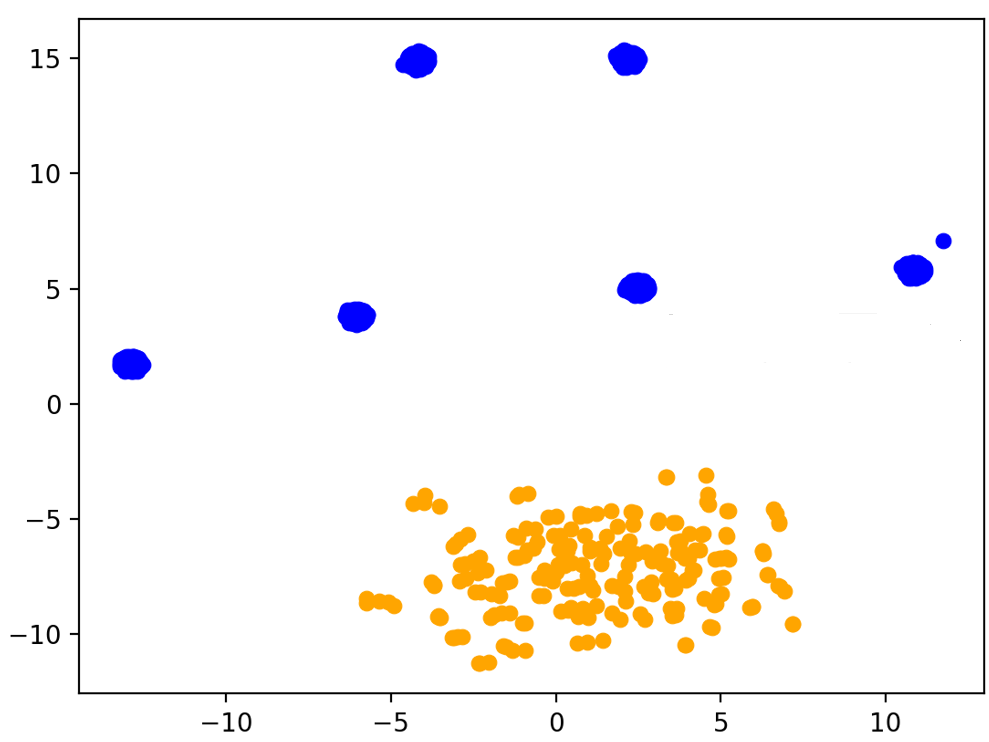}
& \includegraphics[width=0.17\textwidth, height=0.18\textwidth]{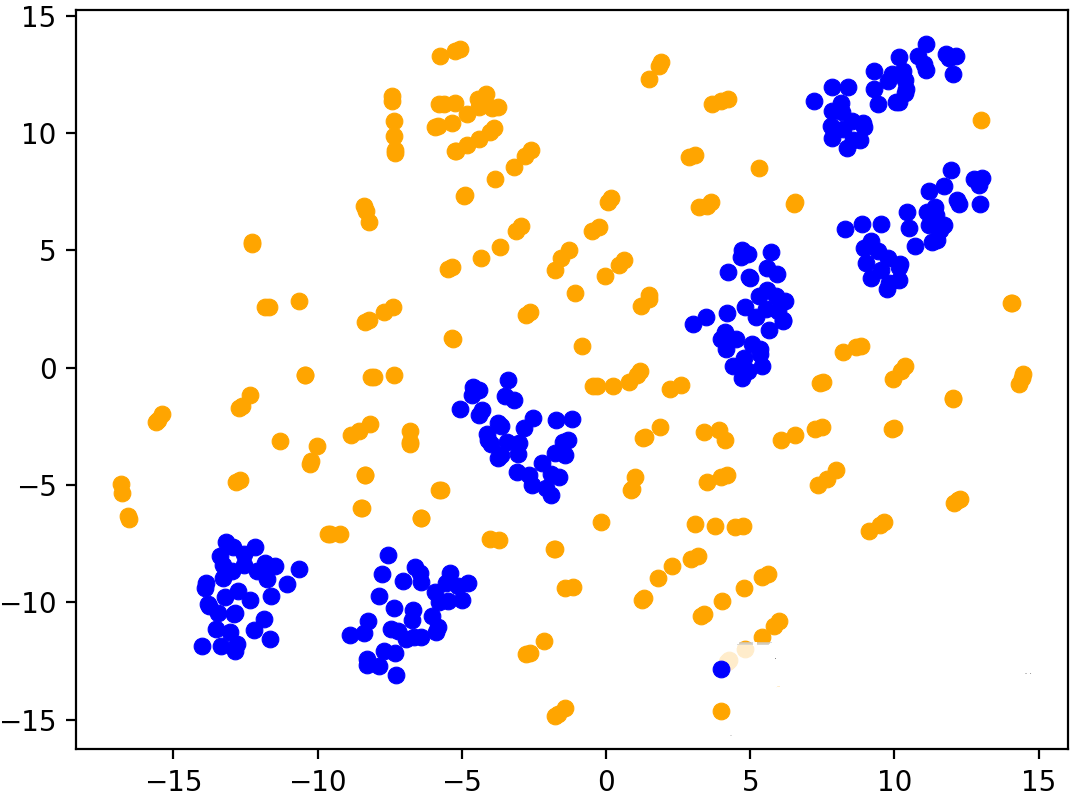}
& \includegraphics[width=0.17\textwidth, height=0.18\textwidth]{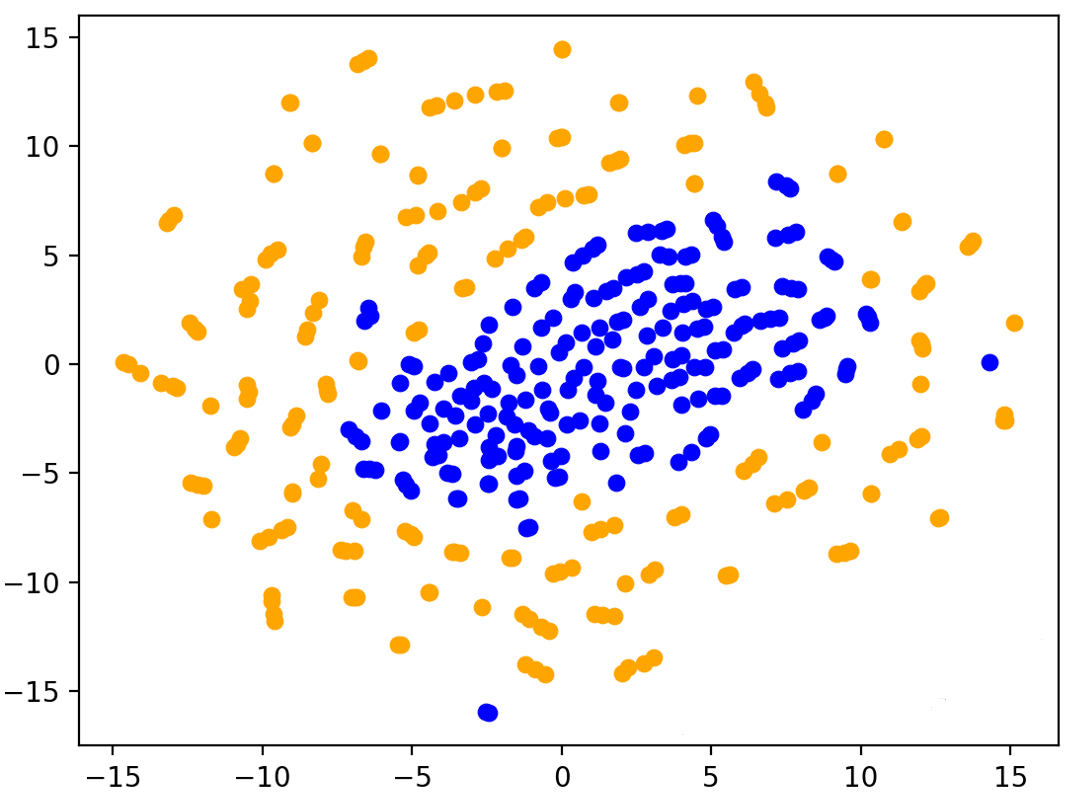}
& \includegraphics[width=0.17\textwidth, height=0.18\textwidth]{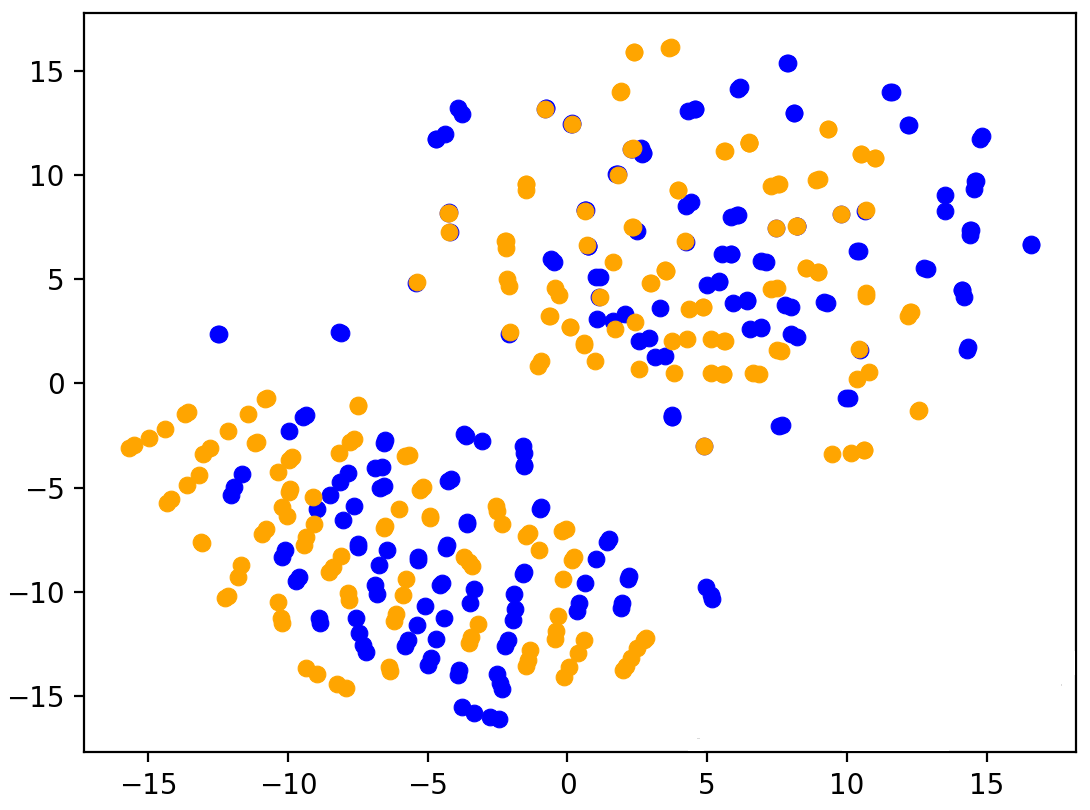} \vspace{0mm}\\
\textbf{\scriptsize{Real}} & {\scriptsize{HA-GAN}} & \scriptsize{3D-$\alpha$-WGAN} &\scriptsize{3D-DPM} & \textbf{\scriptsize{Ours(cDPM)}} \\
\end{tabular}
\caption{\textbf{Left:} One MRI generated by our model and its closest real MRI based on MS-SSIM. \textbf{Right:} tSNE embedding of 200 generated samples (\textcolor{blue}{blue}) of each model and their closest real MRIs (\textcolor{Dandelion}{orange}). Only our model generated independent and diverse samples as the data points overlay but are not identical to the training data.}
\label{fig:tsne}
\end{figure}

\subsection{Results}
\subsubsection{Qualitative results} The center of the axial, coronal, and sagittal views of five MRIs generated by cDPM shown in Fig.~\ref{showcases} look realistic. Compared to the MRIs produced by the other approaches other than 3D-DPM (see Fig.~\ref{fig:comparetoGANs}), the MRIs of cDPM are sharper; specifically, the gray matter boundaries are more distinct and the scan provides greater anatomical details. As expected, 3D-DPM produced synthetic slices of similar quality as cDPM but failed to do so for the entire MRI.

The synthetic MRIs of cDPM shown in Fig.~\ref{showcases} are also substantially different from each other, suggesting that our method could be used to create an augmented data set that is anatomically diverse. Fig.~\ref{fig:tsne} further substantiates the claim, which plots the t-SNE embedding \cite{van2008visualizing} of 200 synthetic MRIs (blue) and their closest real counterpart (orange) according to MS-SSIM for each method. Note, matching of all 500 synthetic MRIs was computationally too expensive to perform (takes days to complete per method). Based on those plots, cDPM is the only approach able to generate MRIs, whose distribution resembled that of the real MRIs. This finding is somewhat surprising given that the MRI subvolumes generated by 3D-DPM looked real. Unlike the real data, however, their distributions are clustered around the average. Thus, 3D-DPM fails to diversify the data set even if (in the future) more computational resources would allow the method to generate a complete 3D MRI.

\begin{table}[!t]
    \caption{Measuring the quality of 500 synthetic MRIs.  \mbox{`( )'} contains the absolute difference to the MS-SSIM score of the real MRIs, which was 0.792. In bold are the optimal scores among methods that generate the entire volume. Scores denoted with an asterisk `*' are only computed on 32 slices.}
 \label{tab1}
    \centering
    \begin{tabular}{l|>{\centering\arraybackslash}p{0.19\textwidth}|>{\centering\arraybackslash}p{0.14\textwidth}|>{\centering\arraybackslash}p{0.14\textwidth}|>{\centering\arraybackslash}p{0.14\textwidth}|>{\centering\arraybackslash}p{0.11\textwidth}}
    
    \hline
      &  MS-SSIM & MMD$\downarrow$ & FID-A $\downarrow$ & FID-C $\downarrow$& FID-S $\downarrow$\\
      & (\%) & (10$^3$) &  &  &  \\
    \hhline{======}
    3D-VAE-GAN~\cite{larsen2016autoencoding} &  88.3 (9.1)&    5.15   &  320      & 247        & 398 \\
    3D-GAN-GP~\cite{gulrajani2017improved} &  81.0 (1.8)&     15.7   & 141     & 127        & 281  \\
    3D-$\alpha$-WGAN~\cite{kwon2019generation} & 82.6 (3.4) & 13.2    & 121     & 116        & 193 \\
    CCE-GAN~\cite{xing2021cycle} &              81.5 (2.3)&   3.54    & 69.4       &  $869$     & 191 \\
    HA-GAN~\cite{sun2022hierarchical}      &             36.8 (42.4)  &  226    & 477      &  $1090$     & 554 \\
    3D-DPM~\cite{dorjsembe2022three} &          79.7 (0.5)* &  15.2*   & 188       &  -    &  - \\
    \textbf{Ours (cDPM)} & \textbf{78.6 (0.6)}&  \textbf{3.14} & \textbf{32.4} & \textbf{45.8} & \textbf{91.1} \\
    \hline
    \end{tabular}

    \end{table}

\subsubsection{Quantitative results} 
Table~\ref{tab1} lists the average scores of MS-SSIM, MMD, and FID for each method. Among all models that generated complete MRI volumes, cDPM performed best. Only the absolute difference between the MS-SSIM score of 3D-DPM and the real MRIs was slightly lower (i.e., 0.005) than the absolute difference for cDPM (i.e., 0.006). This comparison, however, is not fair as the MS-SSIM score for 3D-DPM was only computed on 32 slices. Further supporting this argument is that FID-A (the only score computed for the same slice across all methods) was almost 5 times worse for 3D-DPM than cDPM. 



\section{Conclusion}

We propose a novel conditional DPM (cDPM) for efficiently generating 3D brain MRIs. Starting with random noise, our model can progressively generate MRI slices based on previously generated slices. This conditional scheme enables training the cDPM with limited computational resources and training data. Qualitative and quantitative results demonstrate that the model is able to produce high-fidelity 3D MRIs and outperform popular and recent generative models such as the CCE-GAN and 3D-DPM. Our framework can easily be extended to other imaging modalities and can potentially assist in training deep learning models on a small number of samples.



 \section{Acknowledgement}
 This work was partly supported by funding from the National Institute of Health (MH113406, DA057567, AA021697, AA017347, AA010723, AA005965, and AA028840), the DGIST R\&D program of the Ministry of Science and ICT of KOREA (22-KUJoint-02), Stanford School of Medicine Department of Psychiatry and Behavioral Sciences Faculty Development and Leadership Award, and by the Stanford HAI Google Cloud Credit. 


%

%

\bibliographystyle{splncs04}
\bibliography{cites}

\end{document}